\documentclass[superscriptaddress,aps,pra,twocolumn,amssymb,amsfonts,letterpaper]{revtex4}
\usepackage{amsmath} \usepackage{amsthm} \usepackage{color}
\usepackage{graphics,graphicx}
\usepackage[pdftex]{hyperref}

\def\one{{\mathchoice {\rm 1\mskip-4mu l} {\rm 1\mskip-4mu l} {\rm
1\mskip-4.5mu l} {\rm 1\mskip-5mu l}}}

\usepackage[matrix,frame,arrow]{xy}
\usepackage{amsmath}
\newcommand{\bra}[1]{\left\langle{#1}\right\vert}
\newcommand{\ket}[1]{\left\vert{#1}\right\rangle}
\newcommand{\qw}[1][-1]{\ar @{-} [0,#1]}
\newcommand{\qwx}[1][-1]{\ar @{-} [#1,0]}
\newcommand{\cw}[1][-1]{\ar @{=} [0,#1]}
\newcommand{\cwx}[1][-1]{\ar @{=} [#1,0]}
\newcommand{\gate}[1]{*{\xy *+<.6em>{#1};p\save+LU;+RU **\dir{-}\restore\save+RU;+RD **\dir{-}\restore\save+RD;+LD **\dir{-}\restore\POS+LD;+LU **\dir{-}\endxy} \qw}
\newcommand{\meter}{\gate{\xy *!<0em,1.1em>h\cir<1.1em>{ur_dr},!U-<0em,.4em>;p+<.5em,.9em> **h\dir{-} \POS <-.6em,.4em> *{},<.6em,-.4em> *{} \endxy}}

\newcommand{\control}{*-=-{\bullet}}

\newcommand{\ctrl}[1]{\control \qwx[#1] \qw}
\newcommand{\lstick}[1]{*!R!<.5em,0em>=<0em>{#1}}

\newcommand{\Qcircuit}{\xymatrix @*=<0em>}

\providecommand{\ignore}[1]{}
\newcommand{\MC}[1]{{\color{blue}[Manny: #1]}}
\renewcommand{\MC}[1]{}

\newcommand{\measureCl}[1]{*+[F-:<.9em>]{#1}
\cw}

\def\tr{{\rm tr}\; }
\def\cO{{\cal O}}

\def\cE{{\cal E}}
\def\cR{{\cal R}}
\def\({\left(}
\def\){\right)}
\def\cS{{\cal S}}
\def\cN{{\cal N}}
\def\cH{{\cal H}}
\def\cZ{{\cal Z}}

\def\hc{{\rm h.c.}}
\def\cost{{\rm cost}}

\def\tpsi{{\tilde{\psi}}}
\def\tH{{\tilde H}}
\def\tDelta{{\tilde \Delta}}

\newcommand{\braket}[2]{\langle#1|#2\rangle}
\newcommand{\sket}[1]{| #1 \rangle}
\newcommand{\sbra}[1]{\langle #1 |}
\newcommand{\avg}[1]{\langle #1 \rangle}

\newtheorem{dfn}{Definition}
\newtheorem{thm}{Theorem}
\newtheorem{lem}{Lemma}
\newtheorem{cor}{Corollary}

\begin{document}

\title{
Eigenpath traversal by phase randomization
}

\author{S. Boixo} \affiliation{Institute for Quantum Information,
  California Institute of Technology, Pasadena, CA 91125, USA}
\email{boixo@caltech.edu}

\author{E. Knill} \affiliation{National Institute of Standards and
  Technology, Boulder, CO 80305, USA}

\author{R. D.  Somma}
\affiliation{Perimeter Institute for Theoretical Physics, Waterloo, ON
  N2L 2Y5, Canada }

\date{\today}
\begin{abstract}  
  A computation in adiabatic quantum computing is implemented by
  traversing a path of nondegenerate eigenstates of a continuous
  family of Hamiltonians. We introduce a method that traverses a
  discretized form of the path: At each step we apply the
  instantaneous Hamiltonian for a random time. The resulting
  decoherence approximates a projective measurement onto the desired
  eigenstate, achieving a version of the quantum Zeno effect. If
  negative evolution times can be implemented with constant overhead,
  then the average absolute evolution time required by our method is
  $\cO(L^{2} /\Delta)$ for constant error probability, where $L$ is
  the length of the path of eigenstates and $\Delta$ is the minimum
  spectral gap of the Hamiltonian. The dependence of the cost on
  $\Delta$ is optimal. Making explicit the dependence on the path
  length is useful for cases where $L$ is much less than the general
  bound. The complexity of our method has a logarithmic improvement
  over previous algorithms of this type. The same cost applies to the
  discrete-time case, where a family of unitary operators is given and
  each unitary and its inverse can be used. Restriction to positive
  evolution times incurs an error that decreases exponentially with
  the cost. Applications of this method to unstructured search and
  quantum sampling are considered. In particular, we discuss the
  quantum simulated annealing algorithm for solving combinatorial
  optimization problems. This algorithm provides a quadratic speed-up
  in the gap of the stochastic matrix over its classical counterpart
  implemented via Markov chain Monte Carlo.
\end{abstract}

\maketitle


\section{Introduction and summary of results}
\label{intro}
Quantum algorithms are often described by means of quantum circuits:
the algorithm starts with a well-characterized pure state; a sequence
of elementary (unitary) gates is applied; and a final projective
measurement in a fixed basis extracts the result.  The circuit model
may not be best for describing all quantum information processing
systems. Adiabatic quantum computing (AQC)~\cite{farhi_quantum_2000},
sometimes also called quantum
annealing~\cite{apolloni_numerical_1988,apolloni_quantum_1988,kadowaki_quantum_1998,santoro_theory_2002,quantum_annealing_2005},
has been proposed as an alternative.

In AQC the computation is performed by smoothly changing the
interaction parameters of the Hamiltonian under which the system
evolves. The initial state is a nondegenerate eigenstate of the
Hamiltonian. The adiabatic theorem of quantum mechanics asserts that
if the continuously related eigenstates remain nondegenerate and the
Hamiltonians change sufficiently slowly, then the final state of the
system is close to the continuously related eigenstate of the final
Hamiltonian~\cite{messiah_quantum_1999}. The last step is a standard
projective measurement. AQC is polynomially equivalent to the quantum
circuit model~\cite{aharonov_adiabatic_2007}.

In this paper we give a method for traversing eigenstate paths of
Hamiltonians that differs from AQC by the use of evolution
randomization. The method is based on previous
results~\cite{childs_quantum_2002,aharonov_adiabatic_2003} in which
the evolution of AQC is replaced by a sequence of projective
measurements onto the instantaneous eigenstate of the Hamiltonian with
the phase estimation algorithm~\cite{kitaev_quantum_1995}, which
exploits the quantum Zeno effect. Both AQC and the Zeno-based model
work, in essence, because an effective level decoupling is introduced
in the Hamiltonian eigenbasis by phase cancellation in AQC or
projections in the Zeno case. Our method also implements a version of
the quantum Zeno effect. We choose a discretization of the eigenstate
path and apply the Hamiltonian corresponding to each point for a
random time. The probability distribution over time may be discrete or
continuous. Consequently, the \emph{randomization method} can also be
used in the case where we are given a path of efficiently
implementable unitary operators and an eigenstate of the last operator
on the path is to be prepared. This case occurs in the quantum
simulated annealing (QSA) algorithm constructed
in Ref.~\cite{somma_quantum_2008}. The probability distribution over
evolution times must be chosen so as to cancel unwanted coherences and
simulate the Zeno effect.

The algorithmic complexity of the randomization method is defined as
the average sum of the absolute evolution times for the Hamiltonians
or by the average number of times the unitaries are applied. The
complexity can be bounded in terms of a lower bound $\Delta$ on the
absolute value of the minimum spectral gap of the Hamiltonians or the
minimum phase gap of the unitaries, the length $L$ of the path of the
states (defined below), and the desired maximum error $\epsilon$ of the
final state compared to the target eigenstate. We show that the
complexity is $\cO( {\log(L/\epsilon)^\alpha L^2/(\epsilon\Delta)} )$,
where $\alpha = 0$ if we can evolve backward and forward in time, and
$1$ otherwise. Backward evolution is possible at the same cost by
reversing quantum circuits for the forward evolution, if such
evolution is circuit-based. To achieve this complexity without
additional dependencies, we use a parametrization of the operators
along the path for which the eigenstates move at a rate that is close
to uniform (up to a constant factor). In many cases of interest,
$L\in\cO(1)$ so that the complexity is of order $1/(\epsilon\Delta)$
up to logarithmic factors. The scaling with the gap is optimal and is
better than the $1/\Delta^3$ of rigorous proofs of the adiabatic
theorem~\cite{jansen_adiabatic_2006,lidar_adiabatic_2008,jordan_quantum_2008}.

An advantage of our approach is that the only requirement on the
Hamiltonians or the unitaries along the path is that the length of the
desired eigenstate path is well-defined. A sufficient condition is
that the time derivative of the operators exists. In terms of bounds
on the Hamiltonians and their derivatives, the worst-case bound is of
order $\| \dot H \|^2/(\epsilon\Delta^3)$ up to logarithmic factors.
This bound comes from the inequality $L \le {\|\dot H\|}/{\Delta}$
[see Eq.~\eqref{eq:plb}] and does not depend on a reparametrization.
It is better than the known worst-case bounds associated with the
adiabatic theorem~\cite{jansen_adiabatic_2006} in that it does not
depend on existence of, or bounds on the second derivative of $H$. The
scaling of the bound with the error is worse, in that it can be made
logarithmic for analytic Hamiltonians paths at the cost of less
favorable dependencies on the other parameters of the
problem~\cite{hagedorn_elementary_2002,lidar_adiabatic_2008}.
Logarithmic scaling can also be obtained with the randomization method
provided the final eigenstate's energy is known. To achieve this
scaling, one can use high precision phase estimation to determine
whether the desired eigenstate has been obtained and repeat the
algorithm if not.

\MC{As written, ``exponential'' had to be replaced with
``logarithmic'', and I reworded for clarity. }

Our method is intuitively explained by the quantum Zeno effect.
Suppose that the path of states $\sket{\tilde
\psi(l_1)},\ldots,\sket{\tilde \psi(l_q)}$ satisfies the condition
that for each $j$, $\sket{\tilde \psi(l_{j-1})}$ is sufficiently close
to $\sket{\tilde \psi(l_{j})}$. If we initialize the state
$\sket{\tilde \psi(l_1)}$ and sequentially apply projections onto the
$\sket{\tilde \psi(l_j)}$, we prepare $\sket{\tilde \psi(l_q)}$ with
good probability. We first consider an idealized strategy, where the
projections are replaced by quantum operations of the form
\begin{align}
\label{eq:mje}
M_{l_j}(\rho) = P_{l_j} \rho P_{l_j} + \cE ((\one-P_{l_j}) \rho(\one-P_{l_j})) \;,
\end{align}
where $P_{l_j} = \sket{\tilde \psi(l_{j})}\sbra{\tilde \psi(l_{j})}$
and $\cE$ is an arbitrary quantum operation that may vary from
instance to instance. This can be thought of as a projective
measurement of $\rho$ onto $\sket{\tilde \psi(l_{j})}$ followed by a
process that does not affect $\sket{\tilde \psi(l_{j})}$. The
fundamental effect of $M_{l_j}$ is to remove coherences between
$\sket{\tilde \psi(l_{j})}$ and orthogonal states. It is this
decoherence that induces the quantum Zeno effect by suppressing
transfer of population to orthogonal states. An approximation of this
effect is achieved if we replace the $M_{l_j}$ by random applications of
Hamiltonians or unitaries with $\sket{\tilde \psi(l_{j})}$ as an
eigenstate. We formalize this claim in Sec.~\ref{randomevol},
Thm.~\ref{thm:res}, and give an upper bound for the error in the
approximation in terms of the characteristic function of the
probability distribution underlying the randomization.

We focus on the Hamiltonian-based version of the randomization method.
The analysis for the unitary version is a straightforward
discretization. In the Hamiltonian version, the randomization method
takes as input a continuous path of Hamiltonians $\cH=\{H(s), s \in
[0,1] \}$, and a nondegenerate eigenstate $\ket{\psi(0)}$ of $H(0)$.
The method aims to output the corresponding nondegenerate eigenstate
of $H(1)$, denoted by $\ket{\psi(1)}$, with high fidelity.

We require that the eigenstates $\ket{\psi(s)}$ are nondegenerate with
$\Delta$ a lower bound on the energy gap. If $\ket{\psi(s)}$ is
differentiable (see Appendix~\ref{app:tech} for the more general
case), we can assume without loss of generality that the phases of the
$\ket{\psi(s)}$ are chosen geometrically, so that
$\braket{\partial_s\psi(s)}{\psi(s)} =0$, which gives a path length
\begin{align}
\label{eq:pl} 
L = \int_0^1
  \|\ket{\partial_s \psi(s)}\| d s\;.
\end{align}
The quadratic cost dependence on $L$ comes from a simple Zeno effect
when an ideal decoherence process according to Eq.~(\ref{eq:mje}) is
used. It is probably not fundamental: Coherent versions of the
adiabatic path achieve scalings $\tilde
O(L)$~\cite{abeyesinghe_speed-up_2008}. The dependence of the cost on
$1/\Delta$ is unavoidable for methods with only oracle access to the
Hamiltonian or unitaries. This can be seen intuitively by noting that
we must, in a sense, distinguish between the desired eigenstate and
the others, which requires that we evolve the relative phases
sufficiently far. More rigorously, an asymptotically better dependence
would result in an unstructured search algorithm better than Grover's,
which is known to be impossible. See Sec.~\ref{grover}.

The paper is organized as follows. In Sec.~\ref{randomevol} we explain
how the quantum Zeno effect can be exploited, show how to approximate
projective measurement operations by means of evolution randomization,
and discuss several probability distributions that are useful for
randomization. The randomization method and its complexity 
are analyzed in Sec.~\ref{randommethod}.  In Sec.~\ref{grover} we show 
that the randomization method
provides the expected quadratic quantum speed-up for the unstructured search
problem. In Sec.~\ref{qsa} we describe the QSA to simulate slowly
varying classical Markov chains. In Sec.~\ref{sec:pe} we show
the equivalence of our randomization method with a coherent version of
the quantum Zeno method implemented via the phase estimation
algorithm, and briefly discuss related works. We summarize in
Sec.~\ref{concl}.

\section{Randomized evolutions}
\label{randomevol}
\subsection{Adiabatic quantum computing using the Zeno effect}
\label{sec:zeno}

The quantum Zeno effect is based on the fact that, for a small
displacement $\delta'$, the probability of projecting
$\ket{\psi(s+\delta')}$ onto $\ket{\psi(s)}$ decreases with
$(\delta')^{2}$, while the distance between states is linear in
$\delta'$~\cite{misra_zenos_1977,home_conceptual_1997,nakazato_temporal_1996}. Therefore,
for the path of states $\{\ket{\psi(s)}\}$, the final state
$\ket{\psi(1)}$ can be prepared from the initial state $\ket{\psi(0)}$
with high fidelity by use of a sequence of measurement projections
onto intermediate states $\ket{\psi(s_{1})},\cdots,\ket{\psi(s_{q})}$,
$0 < s_{1}< \cdots < s_{q} =1$. We choose $s_{j}$ so that the fidelity
of the final state with respect to $\ket{\psi(1)}$ is sufficiently
close to unity. It is not necessary to keep track of the measurement
results at intermediate steps, which gives rise to the following definition.
\begin{dfn}
\label{df:pmo}
  A projective-measurement operation onto $\sket{\tpsi(l)}$ is a
  quantum operation of the form
  \begin{align*}
    M_l(\rho) = P_l \rho P_l +\cE ((\one - P_l) \rho (\one - P_l))\;,
  \end{align*}
with $P_l = \sket{\tpsi(l)} \sbra{\tpsi(l)}$ and $\cE$ arbitrary
quantum operations that may vary with $l$.
\end{dfn}

 We assume a monotonically increasing parametrization $s(l)$,
with $l\in [0,L']$, $s(0)=0$ and
$s(L')=1$. We define $\sket{\tpsi(l)}=\ket{\psi(s(l))}$. (Objects with
a tilde correspond to objects in the new parametrization). Later we
consider $s(l)$ so that $L' = L$, the path length of
Eq.~(\ref{eq:pl}). We formulate the Zeno method for quantum state
preparation as
follows~\cite{childs_quantum_2002,aharonov_adiabatic_2003,somma_quantum_2008}:
\begin{lem}[Zeno effect]
\label{lem:zeno}
  Consider a continuous path of states $\{\sket{\tpsi(l)}\}_{l\in[0,L']}$ and assume
  that, for fixed $d$ and all $\delta$,
  \begin{align*}
    | \braket{\tpsi(l)}{\tpsi{(l+\delta)}} |^2 \ge 1-d^2\delta^2\;.
  \end{align*}
 Then the state
$\sket{\tpsi(L')}$ can be prepared from $\sket{\tpsi(0)}$ with
fidelity $p>0$ by $\lceil (L')^2d^2/(1-p) \rceil$ intermediate
projective-measurement operations.
\end{lem}

\proof{
 Divide $[0,L']$ into $q=\lceil (L')^{2}d^2/(1-p) \rceil$ equal segments
and set $\delta=L'/q$. At every point $l_{j}=j \delta$, $1 \le j \le
q$, we perform a projective-measurement operation
onto $\sket{\tpsi(l_{j})}$. The final state is $M_{l_{q}} \circ
M_{l_{q-1}}\circ \cdots \circ M_{l_{1}} (\rho)$, with
$\rho=\sket{\tpsi(0)} \sbra{\tpsi(0)}$. The output fidelity is bounded
as
  \begin{align}
    \tr & [P_{l_{q}}    (M_{l_{q}} \circ  \cdots \circ M_{l_{1}} (\rho))] \ge \| P_{l_{q}} \cdots P_{l_{1}}
    \sket{\tpsi(0)} \|^2 \nonumber \\ 
    {}={}  & \Pi_{j=1}^{q} | \sbra{\tpsi(l_{j})} \tpsi(l_{j-1})\rangle |^2 
    \nonumber \\
    {}\ge{} & (1-d^2\delta^2)^q \ge 
          1- d^2 L'^2/q \ge 1-(1-p) = p \;
  \end{align}
}

From Lemma~\ref{lem:zeno} and assuming a \emph{uniform}
parametrization, defined to satisfy $L(s(l)) = \tilde L(l) = l$,
$d=1$, and $L' = L$ (see Appendix~\ref{app:uniform}), it follows that
the state $\ket{\psi(1)}$ can be obtained with fidelity $p$ starting
from $\ket{\psi(0)}$ with $\cO(L^2/(1-p))$ projective-measurement
operations.

\subsection{Approximating projective-measurement operations through randomized
  evolutions}
  \label{sec:re}
We assume that evolutions under $H(s)$ for time $t$ can be
implemented at a cost linear in $|t| \| H(s) \|$, as in AQC. That is, we do not
take into account the cost of simulating $H(s)$ for small time
intervals. By rescaling $H(s)$ if necessary, we can assume that $\|
H(s) \| \le 1$. Thus, the cost of the randomization method is
determined by the sum of the absolute evolution times. Although we consider
the case where the evolution time $t$ can be negative, one often
restricts $t$ to be nonnegative. This restriction is justified if the
Hamiltonians are physical without a simple time-reversal procedure,
rather than induced by quantum circuits. In the latter case, evolving
for negative $t$ is as efficient as for positive $t$ and can be
realized by reversing the quantum circuits.

We denote by $\Delta(s)$ the spectral gap for the eigenstate
$\ket{\psi(s)}$ of Hamiltonian $H(s)$. The following results also
apply to the unitary case where we are given operators $U(s)$ and
$\Delta(s)$ is the phase gap. In the unitary case the
distributions over time that are used for randomization must be
concentrated at the integers, and correspond to the number of times
the unitaries are applied.

According to Lemma~\ref{lem:zeno}, the Zeno method does not require
that we keep track of intermediate measurement results. Thus, any
purely dephasing mechanism in the instantaneous eigenbasis of $\tH(l)$
implements a version of $M_{l}$. A natural choice for such a
decoherence mechanism is the evolution induced by $\tH(l)$ for a
(unknown) random time. This is the subject of next theorem, where we
bound the residual coherences in terms of the characteristic function
of the random time distribution.

\begin{thm}[Randomized dephasing]
\label{thm:res}
Let $\sket{\tpsi(l)}$ be a nondegenerate eigenstate of $\tH(l)$, and
$\{ \omega_{j} \}$ be the energy differences to the other eigenstates
$\sket{\tpsi_{j}(l)}$. Let $T$ be a random variable associated with
the time of evolution under $\tH(l)$, and $\cR_l^{T}$ the
corresponding quantum operation. Then there exists a quantum operation
$\cE$ such that, for all states $\rho$,
\begin{align*}
  \| (M_l^\cE - \cR_l^T)(\rho) \|_{\tr} \le \epsilon= \sup_{\omega_{j}} \left|
  \Phi(\omega_{j})\right|\;,
\end{align*}
where $M_l^\cE$ is the projective-measurement operation defined in
Definition~\ref{df:pmo} with $\cE$ specified, and $\Phi$ is the
characteristic function of $T$.
\end{thm}

We give the proof in Appendix~\ref{app:dephasing}. It is based on
computing the coherences after the randomized evolution in terms of
the characteristic function of $T$ as
\begin{align}
  \cR_l^{T} (\sket{\tpsi(l)} \sbra{\tpsi_{j}(l)})  = \Phi(\omega_{j}) \sket{\tpsi(l)} \sbra{\tpsi_{j}(l)}\;.
\end{align}

The average cost of randomization is given in terms of the random
variable $T$ as $\avg{|T|}$, the expected value of the absolute evolution time.
If $T$ takes only positive values the average cost is
given by
\begin{align*}
  \avg{T} = -i [\partial_\omega \Phi](0)\;,
\end{align*}
provided it is finite. Note that if $T<0$ is allowed, then the average cost
can be reduced by shifting $T$'s distribution so that $0$ is a median
of $T$.

We can bound the required average cost per step from below by
$\Omega(1/\Delta)$, with $\Delta$ a lower bound on the smallest gap $\inf_{s} |\Delta(s)|$, 
by means of the following theorem:
\begin{thm}
\label{thm:cost}
Let $T$ be a random variable with characteristic function
$\Phi$. Then, for all $\omega$,
  \begin{align*}
    \cost(T) = \avg{|T|} \ge \frac {1-| \Phi(\omega) |}
    {|\omega|}\;.
  \end{align*}
\end{thm}
\proof{
  From the definition of $\Phi$ we obtain
  \begin{align}
      1 - | \Phi(\omega) | &\le |1-\Phi(\omega)| \le \int |
      1- e^{i \omega t}| d\mu(t) \nonumber \\ & \le \int
      |\omega t | d\mu(t) = \avg{|T|} | \omega| \;,
  \end{align}
with $\mu$ the probability distribution of $T$
}

We want to ensure that after the randomized evolution, the remaining
coherences bounded by $|\Phi(\omega)|$ for $|\omega| \ge \Delta$ are
small. Because of Thm.~\ref{thm:cost}, the average absolute evolution
time can be bounded by $\Omega(1/\Delta)$.

\MC{The previous sentence needed a bit of work for clarity.}

If $T<0$ is permitted, the bound of Thm.~\ref{thm:cost} can be
achieved up to a constant factor. See Example~\ref{ex:sinc} and
Lemma~\ref{lem:pos-opt} in Sec.~\ref{sec:examples}. 
For the case where we are given a path of
unitaries and $T$ is restricted to the integers, it suffices to
consider the characteristic function on the interval $[-\pi,\pi]$. The
results of this section are otherwise unchanged.

Repetition of the randomized evolution step decreases the error
exponentially fast in the number of repetitions, as shown by the following
argument. For independent random variables $T_{1}$ and $T_{2}$, the
characteristic function of the sum $T'=T_{1}+T_{2}$ is
\begin{align}
  \Phi'  = \Phi_{1}  \Phi_{2}\;,
\end{align}
with $\Phi_{i}$ the characteristic function of $T_{i}$.
Thus we have the following lemma (the notation is that of Thm.~\ref{thm:res}):
\begin{lem}\label{lem:exp}
Let $T$ be a random variable with characteristic function $\Phi$, and
$\sup_{\omega_{j}} \left| \Phi(\omega_{j})\right| = \epsilon$. Let
$T'$ be the sum of $n$ independent instances of $T$.
Then there exists a quantum operation $\cE$ such
that for all states $\rho$,
\begin{align*}
  \| (M_l^\cE - \cR_l^{T'})(\rho) \|_{\tr} \le  \epsilon^{n}\;.
\end{align*}
\end{lem}

\subsection{Examples of randomized evolutions}
\label{sec:examples}

We consider some examples of randomized evolution steps involving
different time distributions.
\begin{enumerate}
\item\label{ex:simplest} Consider the case where all the orthogonal
 eigenstates to $\sket{\tpsi(l)}$ are degenerate and the spectral gap
 of $\tH(l)$, denoted by $\omega_{1}$, is known. We
 then choose a random variable $T_{\omega_{1}}$ that takes the values $t=0$ or
 $t=\pi/\omega_{1}$, each with probability $1/2$. The average cost is
 $ \pi/(2 \omega_{1})$. The characteristic function for this
 distribution satisfies
  \begin{align}
   | \Phi (\omega) | = \left| \cos\(\frac {\pi \omega}{2 \omega_{1}}
      \) \right| \;.
  \end{align}
  Since $ \Phi (\omega_{1})=0$ , Thm.~\ref{thm:res} implies that the
  projective measurement onto $\sket{\tpsi(l)}$ can be simulated
  exactly with this distribution. The assumptions in this
  example may seem unrealistic, but it provides a basis for the randomization
  method in unstructured search (Sec.~\ref{grover}).
  It is possible to generalize the method to the case where the
  spectrum is known. If there are $k$ distinct absolute eigenvalue
  differences $\omega_j$, the independent sum of $T_{\omega_j}$
  has the property that the characteristic function is identically
  zero on the eigenvalue differences. The average cost is
  $\sum_j {\pi/(2\omega_j)}$. \MC{Added factor of $\pi$, to match above.}

\item\label{ex:sinc} Let $T$'s probability density be proportional to
  $\textrm{sinc}(\lambda t)^4$, $\lambda>0$. The function
  $\textrm{sinc}$ is defined as $\textrm{sinc}(t) = \sin(t)/t$. The
  Fourier transform of $\lambda\textrm{sinc}(\lambda t)/\pi$ is the
  indicator function of the interval $[-\lambda,\lambda]$. The
  characteristic function of $T$ is therefore proportional to the
  four-fold convolution of this indicator function with itself, which
  is continuous and has support $[-4\lambda,4\lambda]$. There is no
  error in approximating the projective-measurement operation by
  randomized evolution if we choose $4\lambda = \Delta$, with $\Delta$
  a lower bound on the minimum gap. The average cost $\avg{|T|}$ is
  proportional to $1/\lambda = \cO(1/\Delta)$. According to
  Thm.~\ref{thm:cost} this is optimal. A possible problem is that the
  tail distribution of $T$ is large: Moments of order greater than $2$
  are unbounded. Lemma~\ref{lem:pos-opt} shows that this can be
  remedied. For the unitary case we modify $T$ by restricting to the
  integers. That is, we set $\textrm{Prob}(T=n)\propto
  \textrm{sinc}(n\lambda)^4$. For $\lambda \leq \pi/4$, the
  restriction of the characteristic function to $[-\pi,\pi]$ is
  unchanged (see Lemma~\ref{lem:discretization}), so for the case where
  the eigenstate path is determined by a path of unitary operators,
  the same average cost of $\cO(1/\Delta)$ is obtained.

  \MC{I rearranged the references to the lemmas for a better flow of
  thoughts.}

  \item\label{ex:uniform} When the eigenstate path is determined by
  unitary operators, a simple choice of $T$ is the uniform
  distribution on integers between $0$ and $Q-1$, where $Q= \lceil 2
  \pi/\Delta \rceil $. If we repeat the randomization step $n$
  times, we can bound the error with respect to the desired projection
  by (Lemma~\ref{lem:exp})
  \begin{align}
  \epsilon=\sup_{\omega_{j}}  \left| \Phi (\omega_{j}) \right |^{n} \le \left| \frac 1 Q
    \frac{1-e^{i \Delta Q}} {1-e^{i\Delta}} \right|^n \le
    \frac 1 {2^n}\;.
  \end{align}
  The average cost is $n (Q-1)/2 \in \cO(n/\Delta)$. To have error
  at most $\epsilon$, the cost is $\cO(\log(1/\epsilon)/\Delta)$.

  If negative $T$ can be used, we can shift $T$ by $-\lfloor Q/2\rfloor$.
  This does not affect the absolute values of the characteristic function
  but reduces the average cost by a constant factor near $1/2$.
  
\item\label{ex:gaussian} 
  If $T$ is unrestricted,
  we can consider $T$ with Gaussian
  distribution $\cN(0,\sigma)$. Note that restricting to $0$-mean Gaussians
  minimizes $\avg{|T|}$ since the mean and the median coincide.
  The absolute value of the characteristic function is 
  \begin{align}
 | \Phi(\omega) |= \exp\( - \frac {\sigma^2 \omega^2} {2} \)\;.
  \end{align}
 The error of the
  randomization step with respect to the desired projection is bounded
  by
  \begin{align}\label{eq:gc}
    \left|   \Phi( \Delta ) \right| = \exp\( - \sigma^2 \Delta^2 /2\)\;.
  \end{align}
  For this distribution, $\avg{|T|} = \sigma\sqrt{2/\pi}$.
  To have error at most $\epsilon$, we need
  $\sigma\geq 2\log(1/\epsilon)^{1/2}/\Delta$. This gives an average
  cost of $\cO[\log(1/\epsilon)^{1/2}/\Delta]$.

  If $T$ must be positive, we can displace the Gaussian by $x>0$ and
  condition on positive outcomes. The error can be estimated as the
  sum of the probability that the Gaussian is negative, which is
  bounded by $e^{-x^2/(2\sigma^2)}$, and the right-hand-side of
  Eq.~(\ref{eq:gc}). The average cost is $\cO(x+\sigma)$. To have error
  at most $\epsilon$, let $\sigma = 2\log(2/\epsilon)^{1/2}/\Delta$ and
  $x=\sqrt{2}\sigma\log(2/\epsilon)^{1/2}$. The average cost is then
  $\cO[\log(1/\epsilon)/\Delta]$.  According to Thm.~\ref{thm:pw} (below) this
  is optimal for positive T.

\item\label{ex:binomial} In the case where $T$ must be supported on
  integers, one can try to approximate the Gaussian by the shifted binomial
  distribution obtained from the sum of $2m$ independent
  $\{-1/2,1/2\}$ mean-$0$ random variables. The absolute value of the
  characteristic function is
  \begin{align}
  |  \Phi(\omega) | =| \cos(\omega/2) |^{2m}\;.
  \end{align}
  This requires $m\in\Theta(\log(1/\epsilon)/\Delta^2))$ to achieve
  error $\epsilon$ in approximating the desired projection.  The average
  cost is then $\cO(\log(1/\epsilon)^{1/2}/\Delta))$. As in
  Example~\ref{ex:gaussian}, we can shift the distribution by
  $\Theta[\log(1/\epsilon)^{1/2}/\Delta]$ and condition on positive
  integers to ensure that $T$ is positive and obtain an average cost
  of $\cO[\log(1/\epsilon)/\Delta]$.

\end{enumerate}

Except for Example~\ref{ex:sinc}, the distributions above do not
achieve the optimal asymptotic cost for unconstrained $T$.
In the case of Example~\ref{ex:sinc}, the probability density determined by
$\textrm{sinc}(\lambda t)^4$ has long tails and unbounded moments.
This is improved by the following lemma.
\begin{lem}\label{lem:pos-opt} There exist
  probability densities $f$ for $T$ that achieve cost $\avg{|T|} =
  \Theta(1/\Delta)$ and error $|\Phi(\omega)| = 0$ for $|\omega| \ge
  \Delta$, and that have bounded moments of all order, i.e. $\langle
  (T-\langle T \rangle)^{n} \rangle < \infty \ \forall \ n \ge 0$.
  \end{lem}
  \MC{Removed a few ``optimal''s, partly because they are redundant,
    partly because I would have had to add an extra modifier or two.}
  We give a constructive proof in Appendix~\ref{app:lem-opt}. For
  these distributions and $\Delta\leq \pi$, discretization does not
  result in an increase in the error, see the next lemma. Note that in
  the discretized case we are only interested in the region of
  eigenphases $[-\pi,\pi]$ and the relevant gap is the eigenphase gap.

\begin{lem}\label{lem:discretization}
  Let $f$ be a probability density whose characteristic function has
  support in $(-\Delta,\Delta)$ with $\Delta\leq \pi$.
  Then the restriction of $f$ to the integers is a well-defined
  probability distribution with $\textrm{prob}(k) = f(k)$ and
  characteristic function $\Phi(\omega) = 0$ for $|\omega| \in
  (\Delta,\pi]$.
\end{lem}
\MC{Updated the lemma in consideration of the revised proof in the
appendix.} We give the proof in Appendix~\ref{app:discretization}.

For positive $T$ and if only a lower bound $\Delta$ on the gap is
known, it is not possible to improve asymptotically over the shifted
and conditioned Gaussian distribution of Example~\ref{ex:gaussian}:
\begin{thm}
\label{thm:pw}
Let $T$ be a positive random variable with characteristic function $\Phi$. Then
$\sup_{|\omega| \geq \Delta}|\Phi(\omega)|\geq e^{-\Delta\avg{T} \pi/ 2}$.
\end{thm}
The proof is in Appendix~\ref{app:pw}.

\section{The randomization method}
\label{randommethod}

The goal of the randomization method is to prepare the nondegenerate
eigenstate $\ket{\psi(1)}$ of $H(1)$ by traversing 
of the path $\ket{\psi(s)}$. This path is determined by the family
$\cH=\{H(s)\}$. Ideally, we choose the uniform parametrization $s(l)$
discussed in Sec.~\ref{sec:zeno} and Appendix~\ref{app:uniform}. Under
such a parametrization the eigenstates $\ket{\psi(s(l))}$ move at a
constant unit rate along the path. Finding the uniform parametrization
is difficult in general. We therefore consider an arbitrary
\emph{subuniform} parametrization $l\in[0,L']\mapsto s(l)$ so that the
rate at which the states move is bounded by unity. Note that $L' \geq
L$, with $L$ the path length. A subuniform parametrization can usually
be obtained from known properties of $H(s)$; see
Lemma~\ref{lem:dpsi-bound} and Eq.~(\ref{eq:subuniform}) below. We
discretize the path using $q \in \cO((L')^{2})$ segments in order to
achieve bounded error. \MC{Some reordering was required.}

The randomization method uses randomized evolutions $\cR_l^{T}$ to
approximate the projective-measurement operations $M_l$ at values
$s(l)$. Here, $l=k\delta$ for $k=1,\ldots,q$, with $q=L'/\delta$ and
$\delta$ sufficiently small. For good asymptotic behavior, we choose
$T$ as in Lemma~\ref{lem:pos-opt} or Example~\ref{ex:sinc}. If $T$
must be positive, we use the shifted and conditioned Gaussian
distribution of Example~\ref{ex:gaussian}. If $T$ must be restricted
to the integers, as in the case of a path $U(s)$ of unitaries, we use
the discretized version of $T$ (Lemma~\ref{lem:discretization}). We
obtain:

\begin{thm}[Randomization method]
\label{thm:rem}
  There are choices of $q$ and $T$ in the randomization method such
  that the method outputs $\ket{\psi(1)}$ starting from $\ket{\psi(0)}$
  with fidelity at least $p$ and average cost
  \begin{align*}
      \cO\( \frac { (L')^2 \left(\log (L'/(1-p))\right)^\alpha} {(1-p) \Delta} \)\;,
  \end{align*}
  where $\alpha=0$ if $T$ can be negative and $\alpha=1$ otherwise.
\end{thm}

\proof{
We choose a step increment $\delta=L'/q$, with $q=\lceil
2(L')^{2}/(1-p) \rceil$. For this choice, Lemma~\ref{lem:zeno}
guarantees that, if we were to implement the projective-measurement
operations exactly, the error in the preparation of $\ket{\psi(1)}$
would be bounded by $(1-p)/2$, because $d \le 1$ for subuniform
parametrizations. We need to choose $T$ such that the additional
contribution to the error due to the differences between the
randomized evolutions and the projective-measurement operations is
also bounded by $(1-p)/2$. Suppose that the error according to
Thm.~\ref{thm:res} is bounded by $\epsilon$. After $r$ steps we have
\begin{eqnarray}
  \left\|(M_{l_{r}} \circ \cdots \circ M_{l_{1}}  - R_{l_{r}}^T \circ \cdots \circ R_{l_{1}}^{T})(\rho) \right\|_{\tr} \hspace*{-2.5in}&& \nonumber\\
  &=& \left\|(M_{l_{r}} \circ \cdots \circ M_{l_{1}}
              - R_{l_{r}}\circ M_{l_{r-1}} \circ \cdots \circ M_{l_1})(\rho)\right.
             \nonumber\\
  && \;\;{}+\left.(R_{l_{r}}^T\circ M_{l_{r-1}} \circ \cdots \circ M_{l_1}
                    - R_{l_{r}}^T\circ \cdots\circ R_{l_{1}}^T)(\rho)\right\|_{\tr}
   \nonumber\\
  &\leq&\left\|(M_{l_{r}}-R_{l_{r}})(\sigma)\right\|_{\tr}\nonumber\\
      &&\;\;{}   + \left\|R_{l_{r}} (M_{l_{r-1}} \circ \cdots \circ M_{l_{1}}  - R_{l_{r-1}}^T \circ \cdots \circ R_{l_{1}}^{T})(\rho) \right\|_{\tr}
   \nonumber\\
  &\leq& \epsilon + (r-1)\epsilon = r\epsilon\;,
\end{eqnarray}
where we used the fact that quantum operations are trace-norm
contracting, and we implicitly applied induction in the last steps.
The desired bound on the error requires $\epsilon \le {(1-p)/(2q)} =
(1-p)^{2}/ (4(L')^{2})$. According to Lemma~\ref{lem:pos-opt} and
Example~\ref{ex:gaussian} this can be achieved at an average cost
$\avg{|T|}$ of $\cO(1/\Delta)$ if $T$ can be negative, and
$\cO(\log(1/\epsilon)/\Delta)$ otherwise. The total cost for the
procedure is $\cO[q\log(q/(1-p))^\alpha/\Delta]$, and substitution of
the value for $q$ yields the claimed bound
}

For differentiable $H(s)$ and eigenstate path $\ket{\psi(s)}$,
we can obtain a subuniform parametrization $s(l)$
 from bounds on the derivative of $H(s)$
and the gaps. For this we need the next lemma.

\begin{lem}\label{lem:dpsi-bound}
  Suppose that $H(s)$ is differentiable and $\{\ket{\psi(s)}\}$ is a
  path of nondegenerate eigenstates of $\{H(s) \}$ with
  spectral gap $\Delta(s) \ne 0$. Then
  \begin{align*}
    \parallel \sket{\partial_{s} \psi(s)} \parallel \le \frac{ \parallel
      \partial_{s} H(s) \parallel}{|\Delta(s)|} \; .
  \end{align*}
\end{lem}
The proof is in Appendix~\ref{app:dpsi-bound}.

Define $\|\dot H \| = \sup_{s} \|\partial_s H(s) \|$. We
obtain
\begin{align}\label{eq:plb}
  L = \int_0^1  \|\ket{\partial_s \psi(s)}\| d s \le L' = \frac {\|\dot    H\|}{\Delta} \;,
\end{align}
with $\Delta$ a lower bound to the minimum absolute value of the  gap. 
This $L'$ is achieved for the parametrization 
\begin{align}
\label{eq:subuniform}
s(l)= \frac \Delta {\|\dot    H\|}  l \; ,
\end{align}
which is subuniform in general. Using this parametrization we obtain the following
corollary:
\begin{cor}
  Let $H(s)$ be a differentiable path of Hamiltonians
 and $\Delta$ a lower bound on the
  minimum absolute value of the spectral gap.
  Then we can prepare $\ket{\psi(1)}$ from $\ket{\psi(0)}$ with
  bounded error probability at cost
    \begin{align*}
    \cO\(\frac {\|\dot H \|^2}{\Delta^3} \left(\log \(\|\dot H\| / \Delta\)\right)^\alpha \)  \;,
  \end{align*}
  where $\alpha = 0$ if we can evolve for negative times and $\alpha=1$
  otherwise.
\end{cor}

To conclude this section we consider the following two questions: What
is the probability that the cost of the randomization method exceeds
the average cost by a constant factor? How does the actual path
followed by the states obtained in a given instance of the
randomization compare to the adiabatic path?

The average cost of the randomization method is $\avg{C}=q\avg{|T|}$,
where $q$ is defined in the proof of Thm.~\ref{thm:rem}, with $T$ the
relevant random variable. The probability $\textrm{prob}(C \ge a\avg{C})$
is therefore at most $1/a$ (Markov's inequality). If the higher-order
moments of $T$ are bounded, better bounds can be obtained. In
particular, for the distributions whose characteristic functions have
smooth, compact support, $\textrm{prob}(C \ge a\avg{C})$ decreases
superpolynomially in $a$. For $T$ based on Gaussians, the decrease is
$e^{-\Omega(a^2)}$. Since $C$ is determined by a sum of $q$
independent instances of $|T|$, better bounds can be obtained for
specific choices of $T$, particularly if $q$ is large. In particular
the variance of $C$ is inversely proportional to $q$ if $T$ has finite
variance and Chebyshev's inequality or, for sufficiently well-behaved
$T$, large-deviation theory can be applied.

A distinguishing feature of the randomization method is that any given
instance involves unitary evolution, which means that the sequence of
states obtained is pure. What is the probability (over the
randomization of the evolution times) that every state in the sequence
of pure states has fidelity at least $1-\gamma$ with respect to the
corresponding eigenstate along the adiabatic path? In view of the
proof of Thm.~\ref{thm:rem}, the probability that the state after the
$r$'th step has fidelity at least $1-kr\epsilon$ with respect to
$\sket{\tpsi(l_r)}$ is at least $1/k$ (by Markov's inequality). In
particular the fidelity of the last state obtained is at least
$1-k(1-p)$ with respect to $\ket{\psi(1)}$ with probability $1/k$. One
can deduce that many of the states obtained in a typical instance of
the randomization method are close to the corresponding states along
the adiabatic path. Given that the deviation from the adiabatic path
executes a kind of random walk, it is reasonable to conjecture that
for appropriate choices of parameters, the probability that all states
obtained are close to the adiabatic path is also high.

\section{Examples of Quantum computations via evolution randomization}

\subsection{Unstructured search}
\label{grover}

In Grover's algorithm~\cite{grover_fast_1996} we want to find a single
marked element $\cS$ in a space of $N=2^n$ elements.  For this, we
build the Hamiltonian
\begin{align}
  H(s)=-[s  \ket \cS \! \bra \cS\; +\; (1-s) \ket +\! \bra + ] \;,
\end{align}
acting on a set of $n$ qubits. Here, $\ket{+}$ is the equal
superposition state and $\ket{\cS}$ the solution state, which is the
computational basis state corresponding to the marked element.
Evolving with $H(s)$ for time $t$ can be done using
$\cO(|t|^{1+\eta})$ conventional oracle calls, with $\eta>0$
arbitrarily small~\cite{cleve_efficient_2008}. For any $s$, $H(s)$ is
nondegenerate in the subspace spanned by $\{\ket{+},\ket{\cS}\}$. If
$\ket{\psi(s)}$ is the eigenstate with largest eigenvalue, we seek to
prepare $\ket{\psi(1)}=\ket{\cS}$ from $\ket{\psi(0)}=\ket{+}$ with
sufficiently high probability. Preparation of $\ket{\psi(1)}$ using
AQC was studied in Ref.~\cite{roland_quantum_2002}.

The energy gap of $H(s)$ can be obtained exactly in the relevant subspace. It is
\begin{equation}
\label{eq:gg}
\Delta(s) = \sqrt{1-4s(1-s) (1-1/N)} \; ,
\end{equation}
which is minimized at $s=1/2$, giving $\Delta=\Delta(1/2)=1/\sqrt N$.
The path length $L$ can also be obtained exactly and, for large $N$,
we have $L \approx \pi /2$ (the states $\ket{+}$ and $\ket{\cS}$
are almost orthogonal). From Thm.~\ref{thm:rem} the average cost of
the randomization method for constant probability of success is
$\cO(1/\Delta) \in \cO(1/\sqrt{N})$ if the parametrization is
uniform. In the large $N$ limit, this parametrization satisfies
\begin{align}
  l(s) \approx \frac 1 2 \arctan\( \frac 1 {\sqrt N} \frac {1-s}{1/2-s}\)\;,
\end{align}
which satisfies $0 \le l(s) \le \pi/2$.
\MC{``angle between adjacent points'' didn't make any sense, and didn't seem
to be needed, given that we are not explaining details here.}
The randomization method then
consists of a sequence of projective-measurements operations at values
\begin{align}
  s_j \approx \frac 1 2 - \frac {\cot(2l_j)}{2 \sqrt N} =    \frac 1 2 - \frac {\cot(2 j \delta)}{2 \sqrt N}
\end{align}
for some $\delta >0$. Note that this is the same evolution path as the
one considered in Ref.~\cite{roland_quantum_2002}, and that the rate of change
of $s$ as a function of $l$ is $\Delta(s(l))$. A possible
choice for $\delta$ is $\pi/4$. At $s=1/2$, we can implement the phase
randomization by evolving under $H(1/2)$ for time $0$ or $\pi/\Delta$,
each with probability $1/2$. This is the distribution in
Example~\ref{ex:simplest} of Sec.~\ref{sec:examples}, and was also used
in Ref.~\cite{childs_quantum_2002}. It outputs the desired state almost
half the time. 

When more than one marked element exist, the above randomization
method can still be used to output a solution with bounded error
probability: the main effect of adding new projectors in $H(s)$ is an
increased spectral gap $\Delta'(s) \ge \Delta(s)$. Thus, the induced
decoherence still simulates an appropriate measurement in the new
eigenbasis. If the uniform distribution is used for the randomization,
then the algorithm is equivalent to the one discussed in
Ref.~\cite{kaye_quantum_2007}, Sec. 8.4.

\subsection{Quantum simulated annealing}
\label{qsa}
As the previous example demonstrates, distinguishing between the cost
induced by the path length and the one induced by the gap has
important advantages, in particular when $L \in \cO(1)$. Without this
distinction, the actual cost of the method can be highly
overestimated. In Ref.~\cite{somma_quantum_2008} we studied quantum
simulations of classical annealing processes via evolution
randomization. An upper bound on the path length in this case is
independent of the minimum spectral gap $\Gamma$ of the classical
Markov chain (i.e., $\Gamma$ is the difference between 1 and the
second largest eigenvalue of the stochastic matrix). Furthermore,
$\Gamma$ can be quadratically increased using Szegedy's quantum
walks~\cite{ambainis_quantum_2007,szegedy_quantum_2004}. For bounded
error probability, the randomization method using these walks has a
cost $\cO(1/\sqrt{\Gamma})$, where we are disregarding the dependency on
other parameters such as error probability and path length. It
provides a quantum speed-up with respect to simulated annealing using
Markov Chain Monte Carlo methods, where the cost is $\cO(1/\Gamma)$.
Quantum state preparation of Gibbs' states using AQC and the Zeno
method was previously studied in Ref.~\cite{aharonov_adiabatic_2003},
but no quantum speed-up was obtained. Recently, a unitary version of
the quantum simulated annealing algorithm (QSA), that uses Grover's
fixed point method, was introduced in
Ref.~\cite{abeyesinghe_speed-up_2008}. The unitary version improves
the dependence of the cost of QSA on output fidelity compared to that
in Ref.~\cite{somma_quantum_2008}. However, the scaling in the gap is
the same.

Basically, QSA is designed to traverse a coherent version of the
classical-state path traversed by classical simulated annealing. The
quantum state path is in a Hilbert space of dimension corresponding to
the size of the classical state space. The classical annealing path we
consider is determined by $\pi_x(\beta) = e^{- \beta E[x]}/\cZ(\beta)$, where
$\pi_{x}$ is the probability of configuration $x$ in the stationary
(Gibbs) distribution.  $E$ is the associated energy or cost function, $\beta$ is
the inverse temperature, and $\cZ(\beta)$ the partition function. The
corresponding path in Hilbert space is given by the \emph{quantum Gibbs
states} $\ket{\psi(\beta)} = \sum_x \sqrt{\pi_x(\beta)} \ket x$. Note
that a measurement in the computational basis samples $x$ with probability
$\pi_{x}(\beta)$. Since
\begin{align}
  \ket{\partial_\beta \psi(\beta)} = \sum_x(\langle E \rangle - E[x]) \sqrt{\pi_x}/2 \ket x\;,
\end{align}
we obtain the following lemma.
\begin{lem}
  For $\beta \in ( 0, \beta_{f})$,
  \begin{align*}
    \| \ket{ \partial_\beta \psi(\beta)} \| = \sigma(\beta)/2\;,
  \end{align*}
 where $\sigma(\beta)$ is the standard deviation of $E$ at inverse temperature $\beta$. 
 The path length
  satisfies $ L \le \beta_f \sigma/2$, with $\sigma = \sup_\beta
  \sigma(\beta)$.

\end{lem}

If $d'$ is the size of the classical state space and $\gamma$ is the
spectral gap of $E$, then the state $\ket{\psi(\beta_{f})}$, for $\beta_f =
\cO((\log d')/\gamma)$, has high probability amplitude in the configuration
that minimizes $E$. With this $\beta_f$, we have
\begin{align}
  L \in \cO\(\frac {\sigma \log d'}{\gamma}\)\;.  
\end{align}
That $L$ is bounded independently of $\Gamma$ is fundamental for the
success of QSA. Using Szegedy's quantum walks we can boost the gap
towards $\cO(\sqrt{\Gamma})$ and achieve the desired cost. The details
of this procedure are explained in
Refs.~\cite{szegedy_quantum_2004,somma_quantum_2008}.

The QSA is basically a sequence of steps, each constructed to prepare
the states $\ket{\psi(\delta )}$, $\ket{\psi(2\delta )}$, $\cdots,
\ket{\psi(\beta_f)}$ from the initial state $\ket{\psi(0)}$,
$\delta \ll 1$. According to Ref.~\cite{somma_quantum_2008},
these states can be prepared by a version of the Zeno effect in which,
at each step, the corresponding Szegedy walk is applied a random
number of times (see Example~\ref{ex:uniform},
Sec.~\ref{sec:examples}). For this distribution the cost of the QSA
is
\begin{align*}  
 \cO\(\frac{L^2}{\sqrt \Gamma} \log L\) \in \cO\( \frac{ \sigma^2 \log^2 d'}{\gamma^2 \sqrt \Gamma} {\log\(\sigma \log d'/\gamma\)} \)\; ,
 \end{align*}
 with $\Gamma$ the minimum gap of the Markov chain along the path. The
 results in Sec.~\ref{sec:examples} show that using the inverses of
 the quantum walk steps, the second logarithmic factor can be dropped.
 Because of the way the quantum walk is constructed, circuits for the
 inverses can be obtained by direct reversal of the circuits for the
 quantum walk steps. In Ref.~\cite{abeyesinghe_speed-up_2008} 
 the authors show that a coherent
 (non-monotonic) path traversal that uses Grover's fixed point method
 for this case can be implemented with an improved 
 cost $\cO(L \log^2L  /\sqrt \Gamma)$.


\section{Relation to other work}
\label{sec:pe}

\begin{figure}[ht]
\begin{align*}
\begin{array}{c}  
\Qcircuit @C=1em
@R=0.75em {
  \lstick{{\ket 0^{\otimes r}}} &{/} \qw & \gate{H^{\otimes r}} &
\raisebox{+.5cm}{{ \mbox{$\ket j$}}} \qw  &\ctrl{1}   & \gate{FT^\dagger} & \meter
& \measureCl{\mbox{Trash}}\\
    \lstick{\rho} &{/} \qw & \qw &\qw 
&\gate{U_l^j} &\qw &\qw&
}\\  \\ \text{(a)} \\ \\ 
\Qcircuit @C=1em @R=0.75em {
    \lstick{{(\one/2)^{\otimes r}}} &{/} \cw & \cw &
\cw &\raisebox{+.3cm}{{ \mbox{$j$}}} \cw  & \control \cw \cwx[1] & & \\
\lstick{\rho} &{/} \qw & \qw &\qw &\qw &\gate{U_{l}} &\qw & &
&
}  \\ \\ \text{(b)} 
\vspace{-.2cm}
\end{array}
\end{align*}
\caption{(a) Phase estimation algorithm. At the end of the algorithm,
  the top $r$-qubit register encodes a $r$-bit approximation to an
  eigenphase of $U_l$ on readout. It is initialized with Hadamard
  gates to an equal superposition state. A sequence of $2^r-1$
  controlled $U_l^j$ operations is applied, and the first register is
  measured after an inverse quantum Fourier transform. If the
  measurement outcome approximates an eigenphase of $U_l$, the second
  register (system) is approximately projected onto the corresponding
  eigenstate. (b) Randomized evolution. If the phase estimation
  algorithm outcome is ignored, the overall effect is equivalent to
  the one induced by initializing a set of $r$ bits (first register)
  in a random state $j$, with $j\in [0..2^r-1]$, and by acting with
  $U_l^j$. Double lines indicate classical information.}
\label{peafig}
\end{figure}
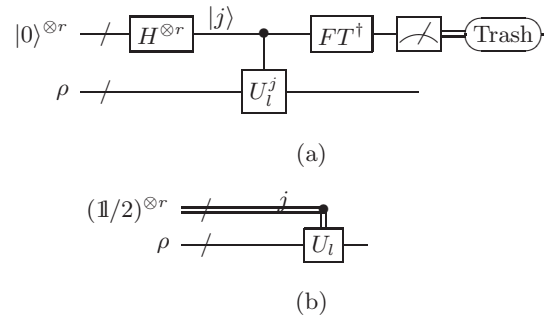

It has been noted
previously~\cite{childs_quantum_2002,aharonov_adiabatic_2003,magniez_search_2006,somma_quantum_2008,abeyesinghe_speed-up_2008}
that the projective-measurement operations $M_l$ can be simulated
using Kitaev's phase estimation algorithm~\cite{kitaev_quantum_1995}
in the discrete-time case. This requires implementing unitaries
$U_l=e^{-i \tH(l)}$ controlled on $r$ ancillary qubits initialized in
the equal superposition state. Then the inverse of the quantum Fourier
transform is applied to the ancillary qubits, and a projective
measurement on the computational basis of the ancillae is performed
[see Fig.~\ref{peafig}(a)]. The phase estimation algorithm needs to
resolve the desired eigenphase from other eigenphases to be able to
project the state of the system into the desired eigenstate. This
requires $2^{r} \in \Omega(1/\tDelta(l))$ uses of controlled-$U_l$'s
for constant error. The error per step has to be small. If one of the
high-confidence versions of the phase estimation
algorithm~\cite{knill_quantum_2006} is used, the overhead to achieve
error $\epsilon$ is logarithmic in $1/\epsilon$. The overall cost is
then similar to that of the randomization method when $T$ is
restricted to be positive.
 
Interestingly, the phase-estimation-based algorithm produces the same
effect on the system as the randomization method if we sample the
evolution time from the uniform distribution on an interval. This is
because the phase estimation ancillary qubits can be traced out after
each step. As a result, the inverse quantum Fourier transform can also be
dropped.  Consequently, the coherence in the state of the ancillary
qubits, initialized in the equal superposition state, plays no role
and these qubits can be replaced by classical bits, each being $0$ or
$1$ with probability $1/2$.
This equivalence was also studied in Ref.~\cite{kaye_quantum_2007}. We
illustrate it in Fig.~\ref{peafig}(b).

Repeating the phase estimation algorithm $n$ times is equivalent to
randomizing with the sum of $n$ independent uniform distributions.
This was considered in Example~\ref{ex:uniform},
Sec.~\ref{sec:examples}. The unwanted coherences reduce exponentially
in $n$.

There are previously noted relationships between the Zeno effect and
coherent evolutions similar to the continuous or discrete evolutions
used in the randomization method. For example, the effect of a strong
interaction with another system, such as might occur in the coupling
to a measurement apparatus, is to restrict the natural Hamiltonian to
the eigenspaces of the interaction~\cite{facchi_quantum_2002}. The
suppression of coherent transitions by randomization with the
interaction Hamiltonian would have a similar effect. A discrete
version of this observation relevant to the analysis of dynamical
decoupling was considered in~\cite{facchi_unification_2004}.

There is a relationship between the way in which interactions are
averaged away in dynamical decoupling, particularly randomized
dynamical decoupling~\cite{viola_random_2005,santos_advantages_2008},
and how transitions between the adiabatic path and the other
eigenstates are suppressed in the randomization method. The
relationship can be made explicit by changing to an $s$-dependent
frame in which the Hamiltonians $H(s)$ are diagonal. In this frame,
the transitions show up explicitly due to the frame changes with $s$.
Strong or randomized evolution under $H(s)$ suppresses these
transitions by averaging them to zero. Dynamical decoupling typically
uses operators that have stronger averaging effects.

A feature of the randomization method is the use of phase decoherence
to ensure a more efficient transfer to a state of physical or
computational interest. There are other ways in which decoherence can
play a role in preparing states for quantum computing. Early examples
proposed the synthesis of {\em decoherence free subspaces} from an
environment-induced quantum Zeno
effect~\cite{beige_quantum_2000,tregenna_quantum_2002,beige_quantum_2003}.
The use of decoherence to decrease the mixing time of quantum walks
was proposed in Ref.~\cite{richter_quantum_2007}. A related phenomenon
has been studied in the context of energy
transfer~\cite{mohseni_environment-assisted_2008,rebentrost_role_2008,rebentrost_environment-assisted_2008,caruso_fundamental_2009}
as realized in certain biological molecules. More generally, it may be
that decoherence or thermal noise can enhance the success probability
in adiabatic quantum computing~\cite{childs_robustness_2001,
  amin_thermally_2008,amin_role_2008,lloyd_robustness_2008}.  Note
that the required thermal noise is different from the
phase-decoherence associated with the randomization method in that it
has the potentially desirable effect of transferring population to
lower-energy eigenstates of the currently active Hamiltonian. Whether
the requirements for effective exploitation of this situation can be
met in realistic devices is not clear. 

Finally, the engineering of a dissipative process using feedback
techniques to stabilize a desired quantum state has been studied
extensively in the quantum control
literature~\cite{wang_feedback-stabilization_2001,wiseman_bayesian_2002,ticozzi_quantum_2008,ticozzi_analysis_2008}. Reference~\cite{verstraete_quantum_2008}
advances that the final state of AQC can be made the unique steady
state of the dissipation, even without feedback. In essence, the
process' Lindblad operators encode the gates of the quantum
computation and appropriate updates to a logical clock
register. Again, the necessary dissipation requires more than the
phase decoherence realized by the randomization method.

\section{Conclusions}
\label{concl}

We have described a method for state preparation in the spirit of AQC,
but based exclusively on randomized evolutions. The idea is to perform
a discrete sequence of projective measurement operations onto the
desired (instantaneous) eigenstate of a given Hamiltonian or unitary
path. These operations are induced via evolution randomization, which
realizes the necessary decoherence in the eigenbasis. We bound the
residual coherences after the randomization in terms of the
characteristic function of the random time. \MC{Changed terms to avoid
having ``measurement'' mean something unconventional.}

We obtained the following exact bounds on the dephasing achieved by
randomized evolutions: First, to induce enough decoherence, the
average evolution time per step scales with the inverse of the minimum
absolute value of the spectral (or eigenphase) gap. Second, repetition
of the randomization reduces the coherences exponentially in the
amount of repetitions. Third, if negative-time evolutions are
implementable with constant overhead, logarithmic factors depending on
the error can be reduced to constant factors, even for discretized
evolutions. \MC{``fundamental scaling'' was not clear enough.
We showed an optimality within the constraints of the method, but
I couldn't think of saying this in a simple clause without having
the reader think it means more than it does. I think that there
is no need say it here.} Fourth, for
non-negative evolutions and if only a lower bound on the absolute
value of the gaps is known, the logarithmic overhead is unavoidable.

We show that the complexity of path traversal algorithms is best
expressed in terms of the path length $L$. The explicit dependence of
the complexity on $L$ can be very helpful when $L$  does not
depend on the gap. This happens, for example, in the Hamiltonian
version of an algorithm for unstructured search, where we showed that
a simple choice of step size and random time distributions rotates
into the solution state with probability $1/2$. One further advantage
of the path-length formulation is that we do not require the
relatively strong differentiability requirements on $H$ as in the
proofs of the adiabatic condition~\cite{jansen_adiabatic_2006} with
explicit bounds as needed for AQC.

Another case where $L$ does not depend on the gap is in the quantum
simulated annealing algorithm, which we also analyzed. This algorithm
provides a quadratic quantum speed-up in terms of the gap with respect
to classical simulated annealing implemented via Markov Chain Monte
Carlo methods. The path is determined by an annealing schedule in
which a parameter $\beta$, related to the inverse temperature of a
classical system, is slowly increased in equal-size steps. The quantum
simulated annealing algorithm allows us to reach the optimal
configuration in time $\cO(1/\sqrt{\Gamma})$ for constant probability
of success and path length, with $\Gamma$ being the minimum gap of the
stochastic matrix (and the corresponding Hamiltonian) along the path.
The improved randomization methods given here remove a logarithmic
factor for the version of the algorithm given
in Ref.~\cite{somma_quantum_2008}.

The similarities of the randomization method with AQC are clear: A
typical instantiation (choice of evolution times) of the randomization
method is, with high probability, an approximation to an adiabatic
path. We find, as is often the case, that it is easier to prove error
bounds for random instances than for the worst case. Whether the
existence statement can be ``derandomized'' efficiently is still an
interesting question. 

\acknowledgments
We thank Howard Barnum for discussions. This work was supported by
Perimeter Institute for Theoretical Physics, by the Government of
Canada through Industry Canada and by the Province of Ontario through
the Ministry of Research and Innovation.  This work was also supported
by the National Science Foundation under grant PHY-0803371 through the
Institute for Quantum Information at the California Institute of
Technology.  Contributions to this work by NIST, an agency of the US
government, are not subject to copyright laws. SB thanks the
Laboratory Directed Research and Development Program at Los Alamos
National Laboratory for support during the initial stages of this
work.

\appendix

\label{app:tech} {\bf Path length.} For states $\ket{\phi_1}$ and
$\ket{\phi_2}$, let $\Theta(\ket{\phi_1},\ket{\phi_2}) =
\arccos(|\braket{\phi_2}{\phi_1}|)$ be the angular distance between
the states. We assume that the $\ket{\psi(s)}$ form a projectively
continuous path, $s\in[0,1]$. The length is given by
\begin{align}
\label{eq:mpl}
  L = \sup_{(s_k)} \sum_k \Theta(\ket{\psi(s_{k+1})},\ket{\psi_{s_k}}),
\end{align}
where the ordered sequences $(s_k)$ subdivide $[0,1]$. Note that the
expression in the limit depends monotonically on the $(s_k)$,
increasing in the refinement order. If $\ket{\psi(s)}$ is differentiable, the
expression in Eq.~(\ref{eq:mpl}) reduces to the one in Eq.~(\ref{eq:pl}). 

\appendix

\label{app:uniform}

{\bf Uniform parametrization.}
Let $\tilde L(l)$ be the length of the path $\sket{\tpsi(l')}$ for
$0\leq l'\leq l$, defined as in Eq.~\eqref{eq:mpl}. Suppose that
$\tilde L(l)$ is Lipschitz continuous so that
$\omega(l_1,l_2)=\sup_{l_1\leq l'<l''\leq l_2} (\tilde L(l'')-\tilde
L(l'))/(l''-l')$ is finite. Note that if $\tilde L$ is
differentiable, one can take $\omega(l_1,l_2) = \sup_{l_1\leq l\leq
l_2} {d\tilde L(l)\over dl}$. In particular, if $\sket{\tpsi(l)}$ is
differentiable, $\omega(l_1,l_2)=\sup_{l_1\leq l\leq l_2} \|
\partial_l\sket{\tpsi(l)} \|$ works [see Eq.~\eqref{eq:pl}]. We
obtain:

\begin{lem}
\label{lem:overlap}
The squared overlap $|\braket{\tpsi(l+\delta)}{\tpsi(l)}|^{2}$ can be
bounded by
  \begin{align*}
    |\braket{\tpsi(l+\delta)}{\tpsi(l)}|^2  \ge 1-
    \omega(l,l+\delta)^2\delta^2 \;.
  \end{align*}
\end{lem}
\proof{
  We have 
  \begin{eqnarray}
    |\braket{\tpsi(l+\delta)}{\tpsi(l)}|^2 & = &
       \cos(\Theta(\sket{\tpsi(l+\delta)},\sket{\tpsi(l)}))^2 \nonumber\\
       &\ge& 1-\Theta(\sket{\tpsi(l+\delta)},\sket{\tpsi(l)})^2 \nonumber\\
       &\ge& 1-(\tilde L(l+\delta)-\tilde L(l))^2 \nonumber\\
       &\ge& 1-\omega(l+\delta,l)^2\delta^2\;
  \end{eqnarray}
}

To take advantage of Lemma~\ref{lem:zeno}, it helps to parametrize the
path with an $s(l)$ for which $\omega(l_1,l_2)$ is as uniform as
possible. For this purpose, define $s(l) = \inf\{s: L(s)\geq l\}$ for
$0\leq l\leq L$, where the length $L(s)$ is the length of the path
$\ket{\psi(s')}$, $0\leq s'\leq s$. The function $s(l)$ is not
necessarily continuous. 

Continuity of states and finiteness of $L$ implies continuity of
$L(s)$. This can be shown as follows: Suppose that $L(s)$ is not
continuous at $s$. Then either $\sup_{\delta > 0}L(s-\delta)<L(s)$ or
$\inf_{\delta > 0}L(s+\delta)>L(s)$. Consider the first case. We have
$L(s) = \limsup_{\delta>0}(\Theta(\ket{\psi(s)},\ket{\psi(s-\delta)})
+ L(s-\delta))$. The inequality implies that $\limsup_{\delta >0}
\Theta(\ket{\psi(s)},\ket{\psi(s-\delta)}) > 0$, contradicting
continuity of $\ket{\psi(s)}$. For the second case, $s<1$. Define
$L(s_1,s_2)$ as the length of the path from $\ket{\psi(s_1)}$ to
$\ket{\psi(s_2)}$. It can be seen from the definition, monotonicity in
the refinement order of the term in the limit of the definition, and
from projective continuity of $\ket{\psi(s)}$ that $L(s,1)=
\sup_{\delta>0}
L(s+\delta,1)+\Theta(\ket{\psi(s)},\ket{\psi(s+\delta)}) =
\sup_{\delta>0}L(s+\delta,1)$ and $L(s,1) =
L(s+\delta,1)+L(s,s+\delta)$. It follows that $\inf_{\delta>0}
L(s,s+\delta) = 0$. The observation now follows from $L(s+\delta)=
L(s)+L(s,s+\delta)$. \MC{Sorry, the second case was not quite
``similar'' enough. Please check that my argument works.}

We define $\tilde L(l)$ as the length of the path $\sket{\tpsi(l')} =
\ket{\psi(s(l'))}$ for $0\leq l'\leq l$. We show that $\tilde L(l) =
L(s(l)) = l$. The second inequality follows from continuity of $L$ and
the definitions. From the definition of path length and since any
subdivision $(l_k)$ of $[0,l]$ corresponds to a subdivision $(s(l_k))$
of $[0,s(l)]$, $\tilde L(l)\leq L(s(l))$. To show the reverse
inequality, let $\bar s = s(L(s))$. Then $\bar s \leq s$ and
$\Theta(\ket{\psi(\bar s)},\ket{\psi(s)}) = 0$. Hence for all
$s'\in[\bar s, s]$, $\ket{\psi(s')} \propto \ket{\psi(\bar s)}$ (that
is, the two states are projectively identical). Consequently, the
right-hand side of Eq.~(\ref{eq:mpl}) is unchanged if we replace the
$s_k$ by $\bar s_k$. Since the $\bar s_k$ are in the range of
$l\mapsto s(l)$, we can choose $l_k=L(\bar s_k)$ to show that the
defining supreme for $\tilde L(l)$ and for $L(s(l))$ are the same.
\MC{This argument needed to be fixed.}

By the previous paragraph, $\omega(l_1,l_2) = 1$ for the
parametrization $s(l)$. We therefore refer to $s(l)$ as the
\emph{uniform} parametrization.

\appendix

{\bf Proof of Theorem~\ref{thm:res}}\label{app:dephasing}
  Let $\mu$ be the probability distribution of $T$.
  For any $\cE$,
   \begin{align} 
     (\cR_l^{T} - &M_l^\cE)(\sket{\tpsi(l)} \sbra{\tpsi_{j}(l)})  \nonumber \\ 
     &=  \cR_l^{T} (\sket{\tpsi(l)} \sbra{\tpsi_{j}(l)})  \nonumber \\ 
     &=  \int e^{-i \tH(l) t} (\sket{\tpsi(l)} \sbra{\tpsi_{j}(l)})  e^{i \tH(l) t} d\mu(t)  \nonumber \\ 
     & =  \int e^{i \omega_j t} d\mu(t)  \sket{\tpsi(l)} \sbra{\tpsi_{j}(l)}\nonumber \\ 
     & = \Phi(\omega_{j}) \sket{\tpsi(l)} \sbra{\tpsi_{j}(l)}\;.
  \end{align}
  We assume without loss of generality that $\rho$ is pure,
  $\rho=\ket{\phi}\!\bra{\phi}$. Write
  \begin{align}
   \ket \phi =  c_{1} \sket{\tpsi(l)} + \sum_{j>1} c_j \sket{\tpsi_{j}(l)} \;.    
  \end{align}
  Let $\cal S$ be the subspace orthogonal to $\sket{\tpsi(l)}$. The
  operation $\cR_l^T$ leaves $\cal S$ invariant, and we can choose
  $\cE=\cR_l^T$ in that subspace. Then
  \begin{align}
  \label{cohbound}
     \big\| (M_l^\cE - &\cR_l^T)(\ket \phi \bra \phi) \big\|_{\tr} \nonumber \\ 
      &= \Bigg\|  \cR_l^{T}  \( \sum_{j>1} c_1 c_j^* \sket{\tpsi(l)} \sbra{\tpsi_{j}(l)}
        + \hc \)  \Bigg\|_\tr  \nonumber \\
     & = \left\| \sum_{j>1} \(\Phi(\omega_{j}) c_1 c_j^*  \sket{\tpsi(l)} \sbra{\tpsi_{j}(l)} + \hc \)\right \|_\tr\;.
  \end{align}
  This is the trace norm of a matrix having 
  \begin{align}
    \pm \sqrt{\sum_{j>1} |\Phi(\omega_{j}) c_1 c_j^*|^2}
  \end{align}
 as the only non-zero eigenvalues. Because of the normalization, 
 $|c_1|^{2}\sum_{j>1} |c_{j}|^{2} \le 1/4$. Thus
 \begin{align}
 \big\| (M_l^\cE - &\cR_l^T)(\ket \phi \bra \phi) \big\|_{\tr} \nonumber \\
   &= 2\sqrt{\sum_{j>1} |\Phi(\omega_{j}) c_1 c_j^*|^2} \nonumber \\
   &\leq \sup_{\omega_{j}} \left| \Phi(\omega_{j})\right| 2 \sqrt{\sum_{j>1} |c_1 c_j^*|^2} \nonumber\\
   &\leq \sup_{\omega_{j}} \left| \Phi(\omega_{j})\right|\;.
 \end{align}

\appendix

{\bf Proof of Lemma~\ref{lem:pos-opt}.}\label{app:lem-opt}
We start with any smooth even function $\hat h$ of compact support in
$(-1/2,1/2)$. \MC{Deleted ``characteristic'' to avoid confusion.} This
implies that its inverse Fourier transform $h$ is real and all its
moments are bounded since
  \begin{align}
|\langle X^{n} \rangle|=\left|\int_{-\infty}^{+\infty} h(x) x^{n} dx \right|= \left| \frac{\partial^{n}\hat{h}(0)}{\partial \omega^{n}} \right| < \infty \ .
    \end{align}
 We define the characteristic function
  $\Phi_{1}$  to be proportional to the convolution of $\hat h$ with itself,
  \begin{align}
    \Phi_{1} (\omega) \propto (\hat h * \hat h)(\omega)\;.    
  \end{align}
  We normalize such that $\Phi_1(0) = 1$. By construction, the inverse
  Fourier transform of $\Phi_{1}$, denoted by $f_{1}$, is positive,
  normalized to $1$, and rapidly decaying, as desired. To accommodate
  arbitrary spectral gaps $\Delta>0$, we rescale the characteristic
  function as $\Phi_\Delta(\omega) = \Phi(\omega/\Delta)$, which has
  support in $(-\Delta,\Delta)$. Its inverse Fourier transform is a
  probability density function $f_{\Delta}(t)=\Delta f_{1}(\Delta t)$.
  The cost of randomization with $f_{\Delta}$ is
  \begin{align}
  \nonumber
  \langle | T| \rangle_{\Delta} &= \int_{-\infty}^{+\infty} |t| f_{\Delta}(t) dt \\
  \nonumber
  &= \Delta \int_{-\infty}^{+\infty} |t| f_{1}(\Delta t) dt \\
&  = \frac{  \langle | T| \rangle_{1}}{\Delta} \; .
  \end{align}
where $ \langle | T| \rangle_{1}$ is the cost of randomization with
$f_{1}$, and is independent of $\Delta$. It follows that $ \langle | T|
\rangle_{\Delta} \in \Theta(1/\Delta)$, which is optimal.

\appendix

{\bf Proof of Lemma~\ref{lem:discretization}.}
\label{app:discretization}
Consider a probability density $f$ with characteristic function $\Phi$
of support in $(-\Delta,\Delta)$, where $\Delta\le \pi$. Consider
$\sum_k\Phi(\omega+2\pi k) = (\Phi*\hat C)(\omega)$ where $\hat
C(\omega) = \sum_k \delta(\omega-2\pi k)$ is a comb. As a
distribution, $\hat C(\omega)$ is the Fourier transform of the comb
$C(t) = \sum_k \delta(t-k)/(2\pi)$. See, for example, Sec.~2.4
of~\cite{porat_1996}. Using the rules for convolution under the inverse
Fourier transform, we find that the distribution $(f\cdot C)(t) =
\sum_k f(k)\delta(t-k)$ has Fourier transform $\Phi*\hat C$. Because
$(\Phi*\hat C)(0)=\Phi(0)=1$, it follows that $f(k)$ is a probability
distribution with the stated properties.

\appendix

{\bf Proof of Theorem~\ref{thm:pw}.}\label{app:pw}
For $\avg{T}$ infinite, there is nothing to prove. So assume $\avg{T}$
is finite, which implies that the characteristic function is
differentiable. Suppose first that $T$ has a square-integrable
probability density $f(t)$. The characteristic function is then a
``Hardy function'' of class $H^{2+}$ as defined
in Ref.~\cite{dym_quantum_1972}, pg.~162. By noting that for $\alpha>0$,
$\omega\mapsto\Phi(\alpha\omega)$ is also Hardy, the proof of Thm.~2
on pg.~166 of Ref.~\cite{dym_quantum_1972} shows that
\begin{align}
  \label{eq:h1}
  \int_{-\infty}^{+\infty}{\log|\Phi(\alpha\gamma)|\over 1+\gamma^2} d\gamma
    \geq \pi\log|\Phi(\alpha i)|\;,
\end{align}
where $\Phi$ has been analytically extended to the upper half plane.
The analytical extension of $\Phi$ is obtained
by using complex $\omega$ in the Fourier transform. Consequently,
$d\Phi(z)/dz$ is the Fourier transform of $t\mapsto itf(t)$, where defined.
In particular, $|d\Phi(z)/dz|$ is bounded by $\avg{T}$ for $z=i\beta$
with $\beta\geq 0$. Since $\Phi(0)=1$,  we have
$\log|\Phi(\alpha i)|\geq \log(1-\alpha\avg{T})$. The integral
of the inequality in Eq.~(\ref{eq:h1}) can be related to
the desired supremum as follows:
\par\begin{tabular}{l}
$\displaystyle
  \int_{-\infty}^{+\infty}{\log|\Phi(\alpha\gamma)|\over 1+\gamma^2} d\gamma\nonumber$
\end{tabular}
\begin{eqnarray}
 &=& \alpha\int_{-\infty}^{+\infty}{\log|\Phi(\gamma)| \over \alpha^2+\gamma^2}
      d\gamma \nonumber \\
 &\leq& 
      \alpha\int_{-\Delta}^{+\Delta}{\log|\Phi(\gamma)| \over \alpha^2+\gamma^2}
      d\gamma \nonumber \\
 &&\;\;{}+\alpha \log(\sup_{|\gamma|\geq\Delta}|\Phi(\gamma)|)
     \int_{|\gamma|\geq\Delta} {1\over\alpha^2+\gamma^2}d\gamma \nonumber\\
\label{eq:h2}
 &\leq&\log(\sup_{|\gamma|\geq\Delta}|\Phi(\gamma)|)(\pi-2\arctan(\Delta/\alpha))
   \;.
\end{eqnarray}
To drop the first summand in the last step we used the fact that
$|\Phi(\gamma)|\leq 1$ because $\Phi$ is the characteristic function
of a probability distribution. We now let $\alpha\rightarrow 0^+$ and
combine with the earlier inequality to get, to first order in $\alpha$,
\begin{equation}
-\pi \alpha \avg{T} \leq 2 \log(\sup_{|\gamma|\geq\Delta}|\Phi(\gamma)|) \alpha/\Delta\;,
\end{equation}
which gives $e^{-\Delta \avg{T} \frac \pi 2} \leq \sup_{|\gamma|\geq\Delta}|\Phi(\gamma)|$.

Now consider arbitrary positive $T$ with $\avg{T}<\infty$, and with
probability distribution $\mu$. Let $S_\delta$ be uniformly
distributed between $0$ and $\delta$. The probability distribution of
$T+S$ has cumulative distribution
$F(x)=\int_0^x\min(1,(x-y)/\delta)d\mu(y)$, which is
differentiable. The corresponding probability density is given by
$\mu([x-\delta,x])/\delta =\int_{x-\delta}^xd\mu(y)/\delta$ and is
square integrable because
\begin{eqnarray}
\int\mu([y-\delta,y])^2 dy &\leq &
  \int\mu([y-\delta,y]) dy \nonumber\\
  &=& \int\int_{y-\delta}^y d\mu(z) dy \nonumber \\
  &=& \int\int_{z}^{z+\delta} dy  d\mu(z) \\
  &=& \int \delta d\mu(z) = \delta \;.
\end{eqnarray}
Thus $T+S$ is subject to the bound of the Theorem. The characteristic
function of $T+S$ is given by $\Phi(\omega) s_\delta(\omega)$, where
$s_\delta(\omega)$ is the characteristic function of $S_\delta$. The
function $s_\delta(\omega)$ converges uniformly to $1$ on bounded
intervals as $\delta\rightarrow 0^+$. It follows that the desired
bound applies to arbitrary positive $T$.

\appendix

{\bf Proof of Lemma~\ref{lem:dpsi-bound}.}\label{app:dpsi-bound}
   Without loss of generality, the phases of $\ket{\psi(s)}$ are
   geometric.
   Because $\Delta(s)>0$ and $H(s)$ is differentiable, 
   it follows that $\ket{\psi(s)}$ is differentiable. From
   the eigenvalue equation
   \begin{align}
     H(s) \sket{\psi(s)} = E(s) \sket{\psi(s)} \; ,
   \end{align} 
   we get
   \begin{align}
     \label{eq:2}
     \nonumber
     \partial_{s} H(s)  \sket{\psi(s)} + H(s) \sket{\partial_{s} \psi(s)} = \\ 
      \partial_{s} E(s)  \sket{\psi(s)} + E(s) \sket{\partial_{s} \psi(s)}\;.
   \end{align}

   Denote by $\ket{\psi_j(s)}$, $ j\in \{ 2,\ldots,d \}$, the $j$-th eigenstate of $H(s)$, orthogonal to $\ket{\psi(s)}$, and with eigenvalue $E_{j}(s)$. We obtain
   \begin{align}
          \braket{\psi_j(s)}{\partial_{s} \psi(s)} = \frac{\bra{\psi_j(s)} \partial_{s} H(s) \sket{\psi(s)}} {E(s)-E_j(s)} \;.
   \end{align}
  Because the path $\ket{\psi(s)}$ is geometric, 
   $\braket{\psi(s)}{\partial_{s} \psi(s)} = 0$ for all $s$. This gives
  \begin{align}
    \nonumber
    &\parallel \sket{\partial_{s} \psi(s)} \parallel^2  = \sum_{j \ge 2} \frac{| \bra{\psi_j(s)} \partial_{s} H(s) \sket{\psi(s)}|^2} {|E(s)-E_j(s)|^2} \\
  &   \le \frac 1 {\Delta(s)^2} \sum_{j\ge 2} \bra{\psi(s)} \partial_{s} H
    \sket{\psi_j(s)}  \bra{\psi_j(s)} \partial_{s} H(s)  \sket{\psi(s)} \nonumber
    \\ 
    \nonumber
    &  \le \frac 1 {\Delta(s)^2}  \bra{\psi(s)} (\partial_{s} H(s))^2  \sket{\psi(s)} 
        \le \frac {\parallel \partial_{s} H(s) \parallel^2}{\Delta(s)^2}\;.
  \end{align}


\end{document}